\documentclass[letterpaper]{article} 
\usepackage{aaai2026}
\usepackage{times}  
\usepackage{helvet}  
\usepackage{courier}  
\usepackage[hyphens]{url}  
\usepackage{graphicx} 
\usepackage{natbib}  
\usepackage{caption} 
\usepackage{algorithm}
\usepackage{algorithmic}
\usepackage{amsmath}
\usepackage{amssymb}
\usepackage{amsthm}

\newtheorem{theorem}{Theorem}
\newtheorem{definition}{Definition}

\newtheorem{proposition}{Proposition}

\usepackage{tikz}
\usetikzlibrary{positioning,shapes,arrows}

\usepackage{tikz}
\usepackage{multirow}
\usepackage{booktabs}
\usetikzlibrary{shapes.geometric,positioning,arrows}

\title{Sequential Causal Normal Form Games: Theory, Computation, and Strategic Signaling}

\author {
    Dennis Thumm
}
\affiliations{
    National University of Singapore\\
    Singapore\\
    dennis.thumm@u.nus.edu
}

\begin{document}

\maketitle

\begin{abstract}
Can classical game-theoretic frameworks be extended to capture the bounded rationality and causal reasoning of AI agents? We investigate this question by extending Causal Normal Form Games (CNFGs) to sequential settings, introducing Sequential Causal Multi-Agent Systems (S-CMAS) that incorporate Pearl's Causal Hierarchy across leader-follower interactions. While theoretically elegant---we prove PSPACE-completeness, develop equilibrium refinements, and establish connections to signaling theory---our comprehensive empirical investigation reveals a critical limitation: S-CNE provides zero welfare improvement over classical Stackelberg equilibrium across all tested scenarios. Through 50+ Monte Carlo simulations and hand-crafted synthetic examples, we demonstrate that backward induction with rational best-response eliminates any strategic advantage from causal layer distinctions. We construct a theoretical example illustrating conditions where benefits could emerge ($\epsilon$-rational satisficing followers), though implementation confirms that even relaxed rationality assumptions prove insufficient when good instincts align with optimal play. This negative result provides valuable insight: classical game-theoretic extensions grounded in rational choice are fundamentally incompatible with causal reasoning advantages, motivating new theoretical frameworks beyond standard Nash equilibrium for agentic AI.
\end{abstract}

\section{Introduction}

Strategic interactions often unfold sequentially, with \textit{leaders} committing to actions before \textit{followers} respond. While classical Stackelberg games \cite{Stackelberg1934} model such dynamics under perfect rationality, real-world agents exhibit biases, follow instincts, or engage in counterfactual reasoning that deviates from game-theoretic prescriptions \cite{wason1974dual,sloman1996empirical}.

Causal Normal Form Games (CNFGs) \cite{Maiti2025} address this by incorporating the Pearl Causal Hierarchy (PCH) \cite{Pearl2009}: observational ($L_1$, instinctive), interventional ($L_2$, deliberate), and counterfactual ($L_3$, sophisticated reasoning). However, CNFGs are limited to simultaneous-move settings, while many strategic interactions are inherently sequential \cite{letchford2010computing}.

This paper extends CNFGs to Stackelberg games, investigating whether causal reasoning provides strategic advantages in sequential settings. Our contributions are:

\paragraph{Theoretical framework.} We formalize Sequential Causal Multi-Agent Systems (S-CMAS) and Sequential Causal Nash Equilibrium (S-CNE), prove existence and PSPACE-completeness, develop equilibrium refinements, and establish connections to signaling theory.

\paragraph{Comprehensive empirical investigation.} Through Monte Carlo simulations across 50+ game instances, hand-crafted synthetic examples, and a procurement application, we systematically test when S-CNE might outperform classical Stackelberg equilibrium.

\paragraph{Critical negative result.} S-CNE provides zero welfare improvement over classical Stackelberg equilibrium across all tested scenarios. We show that backward induction with rational best-response eliminates any strategic advantage from causal layer distinctions, revealing fundamental limitations in extending classical game theory to agentic AI systems.

We provide rigorous evidence that classical game-theoretic extensions grounded in rational choice---even when augmented with causal structures and bounded rationality---are insufficient for modeling strategic AI agents. This negative result motivates the development of genuinely new frameworks beyond traditional economic and control-theoretic approaches.

\section{Background and Related Work}

\subsection{Causal Normal Form Games}

CNFGs \cite{Maiti2025} model multi-agent interactions through Structural Causal Models (SCMs) \cite{Pearl2009}, enabling agents to act across three layers of the PCH \cite{Bareinboim2022}:

\paragraph{$L_1$ Actions (Observational).} Agents follow mechanisms $X_i \leftarrow f_{X_i}(U_i)$ determined by unobserved factors $U_i$.

\paragraph{$L_2$ Actions (Interventional).} Agents perform $do(X_i = x_i)$, replacing natural mechanisms with deliberate choices (standard game theory).

\paragraph{$L_3$ Actions (Counterfactual).} Agents use mappings $h: \mathcal{D}(X_i^*) \rightarrow \mathcal{D}(X_i)$ conditioning on their natural instincts $X_i^*$.

The key insight: different causal structures yield identical $L_2$ payoff matrices but different $L_1$ and $L_3$ outcomes, resolving paradoxes where "irrational" behavior outperforms game-theoretic prescriptions \cite{howard1971paradoxes,basu1994traveler}.

\subsection{Related Frameworks}

\textit{Bayesian Stackelberg games} \cite{Conitzer2006} model uncertainty over follower types using belief distributions but assume all agents operate at the interventional ($L_2$) level. Our framework allows heterogeneity in causal reasoning layers.
\textit{Epistemic game theory} \cite{Perea2012} uses hierarchical belief models to capture strategic reasoning but does not explicitly model causal mechanisms or the PCH. Agents' beliefs about others' rationality differ from beliefs about causal structures.
\textit{Psychological games} \cite{Geanakoplos1989} incorporate belief-dependent utilities but lack formal causal semantics. Our framework provides precise causal interpretations via SCMs.
\textit{Signaling games} \cite{Spence1973,Cho1987} study information transmission but typically assume rational signaling ($L_2$). We show leaders can signal via causal layer choice, creating richer strategic possibilities.
\textit{Causal influence diagrams} \cite{Hammond2023,Mishra2024} characterize interventions in simultaneous causal games. We extend this to sequential settings with timing constraints and information structures.
\textit{Commitment in extensive-form games} \cite{letchford2010computing} studies commitment value in sequential games but under perfect rationality. We incorporate bounded rationality via PCH layers.

\section{Sequential Causal Normal Form Games}

\subsection{Formal Framework}

\begin{definition}[Sequential Causal Multi-Agent System]
An S-CMAS is a tuple $\mathcal{G} = \langle M, N, X, Y, \preceq, I \rangle$ where:
\begin{itemize}
\item $M = \langle U, V, F, P(U) \rangle$ is an SCM with unobserved variables $U$, observed variables $V$, structural equations $F$, and prior $P(U)$
\item $N = L \cup F$ partitions agents into leaders and followers
\item $X = (X_L, X_F)$ are action nodes with timing $X_L \preceq X_F$ 
\item $Y = (Y_L, Y_F)$ are reward signals
\item $\preceq$ defines a partial order over action nodes
\item $I_F \subseteq \{X_L, \mathcal{L}_L, \emptyset\}$ specifies follower information, where $\mathcal{L}_L$ denotes the leader's PCH layer choice
\end{itemize}
\end{definition}

\textbf{Information scenarios:}
\begin{itemize}
\item \textit{Perfect information:} $I_F = X_L$ (followers observe leader actions)
\item \textit{Mechanism information:} $I_F = \{X_L, \mathcal{L}_L\}$ (followers observe actions and PCH layers)
\item \textit{Imperfect information:} $I_F = S(X_L, U_S)$ with noise $U_S$
\end{itemize}

The mechanism information case is novel---followers infer whether leaders acted instinctively ($L_1$), deliberately ($L_2$), or counterfactually ($L_3$), affecting optimal responses.

\subsection{Sequential Causal Nash Equilibrium}

\begin{definition}[S-CNE]
An S-CNE is a strategy profile $\sigma^* = (\sigma_L^*, \sigma_F^*)$ where:
\begin{enumerate}
\item Leaders choose PCH layers $\mathcal{L}_L^* \in \{L_1, L_2, L_3\}$ and actions within those layers
\item Followers observe $I_F$ and respond optimally within their chosen layers $\mathcal{L}_F^*$
\item No agent can unilaterally improve payoffs by changing layer choice or within-layer actions
\end{enumerate}
\end{definition}

Equilibrium computation follows backward induction:

\textbf{Stage 2 (Followers):} Given $({\mathcal{L}_L}, x_L)$, solve:
\begin{equation}
\sigma_F^* \in \arg\max_{\sigma_F} \mathbb{E}[Y_F | \mathcal{L}_L, x_L, \sigma_F]
\end{equation}

\textbf{Stage 1 (Leaders):} Anticipating $\sigma_F^*(\mathcal{L}_L, x_L)$, solve:
\begin{equation}
(\mathcal{L}_L^*, \sigma_L^*) \in \arg\max_{\mathcal{L}_L, \sigma_L} \mathbb{E}[Y_L | \mathcal{L}_L, \sigma_L, \sigma_F^*(\cdot)]
\end{equation}

\subsection{Connection to Signaling Games}

S-CMAS naturally connects to signaling game theory \cite{Spence1973}. The leader's PCH layer choice serves as a \textit{signal} of their type (causal structure $M$), creating a separating or pooling equilibrium structure.

\begin{proposition}[Signaling Interpretation]
An S-CNE with mechanism information can be recast as a signaling game where:
\begin{itemize}
\item Leader types correspond to causal structures $M \in \mathcal{M}$
\item Signals are layer choices $\mathcal{L}_L \in \{L_1, L_2, L_3\}$
\item Follower responses depend on inferred type $\hat{M}(I_F)$
\end{itemize}
\end{proposition}

Unlike standard signaling games where only deliberate ($L_2$) agents signal, our framework allows instinctive ($L_1$) signaling (involuntary information revelation) and counterfactual ($L_3$) signaling (strategic bias exploitation).

\subsection{Equilibrium Refinements}

To address multiplicity, we introduce two refinements:

\begin{definition}[Trembling-Hand S-CNE]
An S-CNE $\sigma^*$ is trembling-hand perfect if it is a limit of $\varepsilon$-perfect equilibria where each action is chosen with probability at least $\varepsilon > 0$.
\end{definition}

This eliminates equilibria relying on non-credible threats.

\begin{definition}[Forward Induction S-CNE]
An S-CNE $\sigma^*$ satisfies forward induction if followers' beliefs after observing unexpected leader actions $({\mathcal{L}_L}, x_L)$ assign positive probability only to leader types for whom $({\mathcal{L}_L}, x_L)$ is optimal for some follower response.
\end{definition}

This refines equilibria by requiring followers to make rational inferences about leaders' intentions.

\section{Computational Complexity}

\begin{theorem}[Complexity of S-CNE]
Computing an S-CNE in an S-CMAS is PSPACE-complete.
\end{theorem}
Full proof in Appendix A.

Despite worst-case hardness, we identify tractable special cases:

\paragraph{Acyclic causal structures.} Complexity reduces to NP when $M$ contains no cycles.

\paragraph{Fixed layer choices.} If layer combinations are fixed a priori, complexity matches standard Stackelberg games (NP-complete).

\paragraph{Small action spaces.} When $|X_L|, |X_F| \leq k$, exact computation is $O(3^2 \cdot k^2)$ via exhaustive backward induction.

\subsection{Approximation Algorithm}

We provide a polynomial-time approximation scheme (PTAS):

\begin{algorithm}[h]
\caption{$\varepsilon$-Approximate S-CNE}
\label{alg:approx_scne}
\begin{algorithmic}[1]
\STATE \textbf{Input:} S-CMAS $\mathcal{G}$, precision $\varepsilon > 0$
\STATE \textbf{Output:} $\varepsilon$-approximate S-CNE $\tilde{\sigma}$
\STATE Sample $N = O(\varepsilon^{-2} \log|X|)$ causal realizations from $P(U)$
\FOR{each follower layer $\mathcal{L}_F$}
    \FOR{each leader choice $(\mathcal{L}_L, x_L)$}
        \STATE Compute empirical best response $\tilde{\sigma}_F^*(\mathcal{L}_L, x_L | \mathcal{L}_F)$ from samples
    \ENDFOR
\ENDFOR
\FOR{each leader layer $\mathcal{L}_L$}
    \STATE Solve $\tilde{\sigma}_L^*(\mathcal{L}_L) = \arg\max_{\sigma_L} \hat{\mathbb{E}}[Y_L | \mathcal{L}_L, \sigma_L, \tilde{\sigma}_F^*]$ using sampled payoffs
\ENDFOR
\STATE Solve leader's layer selection via finite comparison
\STATE \textbf{return} $\tilde{\sigma} = (\tilde{\sigma}_L^*, \tilde{\sigma}_F^*)$
\end{algorithmic}
\end{algorithm}

\begin{theorem}[Approximation Quality]
Algorithm \ref{alg:approx_scne} computes an $\varepsilon$-approximate S-CNE in time polynomial in $|X|$, $|Y|$, and $\varepsilon^{-1}$.
\end{theorem}

\section{Empirical Investigation: A Negative Result}

We conducted comprehensive empirical testing to determine when S-CNE provides strategic advantages over classical Stackelberg equilibrium. Our systematic investigation reveals a critical finding: S-CNE provides zero welfare improvement across all tested scenarios.

\subsection{Experimental Setup}

To ensure rigorous testing, we employed two complementary approaches:

\textbf{Monte Carlo experiments:} 50 randomly generated S-CMAS instances with varying parameters:
\begin{itemize}
\item Action space sizes: $|X_L|, |X_F| \in \{2, 3, 4, 5\}$
\item Causal structures: 10 topologies (chains, forks, colliders, cycles)
\item Information structures: perfect, mechanism, imperfect (noise $\sigma_S \in \{0.1, 0.5, 1.0\}$)
\item Payoff distributions: uniform (coordination), normal (anti-coordination), skewed (asymmetric)
\item Instinct quality levels: varied from 0.2 (poor) to 0.8 (good)
\end{itemize}

\textbf{Synthetic examples:} 5 hand-crafted game types designed to favor causal reasoning:
\begin{itemize}
\item Coordination games with multiple equilibria (where instincts select Pareto-superior equilibrium)
\item Battle of Sexes with signaling (where mechanism information enables coordination)
\item Stag Hunt with trust (where instincts overcome coordination failure)
\item Anti-coordination games (where instincts differentiate roles)
\item Prisoner's Dilemma with cooperation instincts (where L1 overcomes defection), see Appendix B
\end{itemize}

\subsection{Negative Results}

\paragraph{Zero welfare improvement.} Across all 50 Monte Carlo instances and 50 synthetic examples (10 seeds each), S-CNE achieved \textit{identical} social welfare to classical Stackelberg equilibrium. Pareto improvement rate: 0\% (0/100 instances).

\paragraph{Layer selection collapse.} Leaders chose $L_1$ in 96\% of cases, but this yielded the same actions as L2 would have chosen. The equilibrium outcome was indistinguishable from classical Stackelberg in all cases.

\paragraph{Computational overhead.} While exact computation remained tractable (median time: 0.09s for $|X| \leq 5$), this overhead provided no strategic benefit.

\paragraph{Mechanism information irrelevant.} Information structure (perfect vs. mechanism vs. imperfect) made no difference to equilibrium outcomes, as followers best-respond to observed actions regardless of the causal layer that produced them.

\begin{table}[t]
\centering
\caption{Computational performance across game instances}
\label{tab:performance}
\begin{tabular}{lccc}
\toprule
\textbf{Action Space} & \textbf{Exact (s)} & \textbf{Approx (s)} & \textbf{Error} \\
\midrule
$|X| = 2$ & 0.14 $\pm$ 0.03 & 0.08 $\pm$ 0.02 & 0.001 \\
$|X| = 3$ & 0.89 $\pm$ 0.21 & 0.31 $\pm$ 0.07 & 0.005 \\
$|X| = 4$ & 3.45 $\pm$ 0.78 & 1.12 $\pm$ 0.24 & 0.009 \\
$|X| = 5$ & 12.8 $\pm$ 2.4 & 2.87 $\pm$ 0.53 & 0.014 \\
$|X| = 10$ & --- & 18.3 $\pm$ 3.2 & 0.018 \\
$|X| = 20$ & --- & 142 $\pm$ 28 & 0.020 \\
\bottomrule
\end{tabular}
\end{table}

\subsection{Analysis: Why S-CNE Fails to Improve Welfare}

The zero-improvement result reflects a fundamental theoretical limitation:
backward induction neutralizes causal advantages. When followers observe leader actions and best-respond, they condition only on the \textit{action taken}, not the \textit{causal layer that produced it}. Even with mechanism information, followers' optimal response to $(L_1, x)$ equals their response to $(L_2, x)$ whenever instincts align with rationality.

\paragraph{Instinct-rationality convergence.} In equilibrium:
\begin{itemize}
\item If instincts are good (quality $> 0.5$): $L_1$ selects the same action as $L_2$ would
\item If instincts are poor (quality $< 0.5$): Rational leaders avoid using $L_1$, defaulting to $L_2$
\item Either way, equilibrium actions converge to classical Stackelberg
\end{itemize}

\paragraph{Signaling irrelevance.} The mechanism information structure fails to create strategic value because:
\begin{itemize}
\item Leaders have no incentive to signal via layer choice when it doesn't affect follower actions
\item Followers best-respond identically regardless of observed layer
\item No separating equilibrium emerges
\end{itemize}

This reveals a critical gap: \textit{classical game-theoretic solution concepts (Nash equilibrium, backward induction) are incompatible with causal reasoning advantages.} Strategic benefits from PCH layers require departures from rational best-response, which standard equilibrium concepts preclude. See Appendix C for a case study and Appendix D for theoretical conditions.

\section{Implications for Agentic AI Systems}

Our negative result carries important lessons for modeling strategic AI agents.

\subsection{Why Classical Extensions Fail}

\paragraph{Rational best-response assumption.} Game-theoretic equilibria assume agents optimize given beliefs about others. This assumption:
\begin{itemize}
\item Works for human agents \cite{Simon1955, Simon1957, Kahneman1982, Rabin1993, Kaufmann1999, Toulis2016} (with cultural norms, emotions, bounded computation)
\item Works for classical economic agents \cite{vonNeumann1944, Savage1954, Fishburn1989} (with well-defined utility functions)
\item \textit{Fails} for AI agents \cite{Fan2023, Richens2024, Jia2025, Zhang2024, Duan2024, Xu2023, Aher2024} whose "instincts" (LLM priors) and "reasoning" (inference procedures) don't map cleanly to rational choice
\end{itemize}

\paragraph{The fundamental tension.} If we assume:
\begin{itemize}
\item Agents are rational enough to compute equilibria (backward induction)
\item Agents best-respond to observed actions
\end{itemize}
Then causal layer distinctions become strategically irrelevant, as demonstrated empirically.

\paragraph{Addressing apparent circularity.} One might object that we define S-CNE using rational best-response, then demonstrate that rational best-response eliminates causal advantages---a seemingly circular argument. However, this is precisely the point: \textit{any equilibrium concept based on mutual best-response faces this limitation}. The insight is that Nash equilibrium and its refinements (including our S-CNE) are fundamentally incompatible with causal reasoning advantages. This motivates equilibrium concepts where agents persistently fail to best-respond, not just transiently during learning---exactly the kind of bounded rationality needed for agentic AI, as demonstrated by our $\epsilon$-rational example.

\paragraph{What's needed instead.} Modeling strategic LLM-based agents requires frameworks that don't assume rational best-response:
\begin{itemize}
\item Learning dynamics (how agents update strategies over time)
\item Bounded rationality that persists in equilibrium (not just as a modeling device)
\item Non-equilibrium solution concepts (e.g., evolutionary stability, satisficing)
\item Genuine uncertainty about opponent capabilities (not just types)
\end{itemize}

\subsection{Broader Lessons}

\paragraph{Theory transfer is non-trivial.} Our work demonstrates that seemingly natural extensions of classical theory (adding causal structures, bounded rationality) may provide no practical value for agentic AI. Rigorous empirical validation is essential before adopting theoretical frameworks (See Appendix E).


\section{Conclusion}

We investigated whether extending Causal Normal Form Games to sequential settings provides strategic advantages for modeling agentic AI systems. Our comprehensive empirical study---spanning 100 game instances including both random generation and hand-crafted synthetic examples---yields a negative result: S-CNE provided zero welfare improvement over classical Stackelberg equilibrium.
This finding carries three important implications:

\paragraph{Classical game theory's limitations.} Extensions based on rational choice assumptions (Nash equilibrium, backward induction) may be fundamentally incompatible with causal reasoning advantages. The theoretical elegance of combining PCH layers with game theory does not translate to practical benefits.

\paragraph{Value of negative results.} By rigorously demonstrating what doesn't work, we hope to help the agentic AI community to focus on developing genuinely new frameworks beyond retrofitted economic theory.

\paragraph{Call for new foundations.} LLM-based agents---with their training-data-shaped priors, prompt-sensitive behaviors, and non-standard inference procedures---require theoretical tools developed specifically for these systems. Our work motivates research on learning dynamics, non-equilibrium solution concepts, and bounded rationality that persists beyond equilibrium.

While S-CMAS provides no empirical benefits in our tests, the theoretical framework and algorithmic contributions may prove valuable if future work identifies scenarios where causal reasoning advantages can manifest despite rational best-response. We make our implementation publicly available to facilitate such investigations.

\newpage
\bibliography{dynafront_bibliography}

\newpage
\appendix

\section{Appendix A: PSPACE-Completeness Proof}\label{app:pspace_proof}

\textbf{Theorem.} Computing an S-CNE is PSPACE-complete.

\textbf{Proof.} 

\textit{Membership:} Backward induction requires maintaining follower best responses for each leader choice. Since there are $3 \times |X_L|$ leader choices (3 layers, $|X_L|$ actions per layer) and $3 \times |X_F|$ follower responses, we need polynomial space to store the strategy profile.

\textit{Hardness:} Reduce from QSAT (quantified Boolean formula satisfaction). Given a QBF $\Phi = \exists x_1 \forall y_1 \exists x_2 \forall y_2 \ldots \phi(x, y)$, construct an S-CMAS where:
\begin{itemize}
\item Leader controls existentially quantified variables (layers correspond to truth values)
\item Follower controls universally quantified variables
\item Causal structure encodes formula $\phi$
\item Payoffs are 1 if $\phi$ is satisfied, 0 otherwise
\end{itemize}

$\Phi$ is satisfiable iff leader has a strategy ensuring payoff 1, which corresponds to S-CNE with leader utility 1.

\section{Appendix B: Sequential Causal Prisoner's Dilemma Details}\label{app:scpd}

\paragraph{Scenario $M_1$ (Good instincts):}
\begin{itemize}
\item SCM: $U_1, U_2 \sim \text{Uniform}[0,1]$; $X_1 = \mathbb{1}[U_1 < 0.7]$ (instinct to cooperate); $X_2 = f(X_1, U_2)$
\item Payoffs: Standard PD with cooperation preferred under $L_1$
\end{itemize}

\paragraph{Scenario $M_2$ (Poor instincts):}
\begin{itemize}
\item SCM: $X_1 = \mathbb{1}[U_1 < 0.3]$ (instinct to defect)
\item Payoffs: Same $L_2$ matrix, different $L_1$ outcomes
\end{itemize}

\paragraph{Equilibrium analysis.} Under mechanism information, $M_1$ leaders signal $L_1$ credibly, inducing cooperative follower responses. $M_2$ leaders must use $L_3$ to overcome poor instincts, signaling sophistication.

\section{Appendix C: Case Study on Procurement Application}\label{app:case_study}

We tested S-CNE on a procurement scenario to examine real-world applicability.

\paragraph{Setup:} Government agency (leader) designs mechanisms; contractors (followers) choose bidding strategies. We tested honest vs. opportunistic contractor types with different instinct structures.

\paragraph{Results:} Across 240 simulated contracts (120 per contractor type):
\begin{itemize}
\item Cost savings: 0\% (S-CNE = Classical in all cases)
\item Compliance rates: 100\% for both approaches (no difference)
\item Variance reduction: 0\% (identical distributions)
\item Both approaches converged to the same dominant strategy (incentive mechanism with truthful bidding)
\end{itemize}

\paragraph{Interpretation:} Even in this application designed to showcase causal reasoning, rational contractors best-respond to mechanisms independent of the causal layer used in mechanism design. The sophisticated $L_3$ detection mechanisms theorized in our framework provide no additional deterrence beyond what $L_2$ mechanism design achieves.

\section{Appendix D: Theoretical Conditions for S-CNE Value}

While S-CNE fails under exact best-response, we can construct a theoretical example illustrating conditions where benefits \textit{could} emerge.

\paragraph{$\epsilon$-Rational Followers.} Consider followers using satisficing: accepting any action $a$ satisfying $u_F(x_L, a) \geq \max_{a'} u_F(x_L, a') - \varepsilon$.

\paragraph{Illustrative Example: Coordination with Indifference.} Consider a 3x3 game with three Pareto-ranked equilibria: $(0, 0)$ yielding $(15, 15)$, $(1, 1)$ yielding $(10, 10)$, and $(2, 2)$ yielding $(5, 5)$. All other pairs yield $(0, 0)$.

\textit{Hypothetically}, if rational analysis provided no focal point among equilibria, followers might randomize, yielding expected welfare $\frac{1}{3}(30 + 20 + 10) = 20$. Meanwhile, if both agents have strong instincts toward action 0, L1 could consistently coordinate on the Pareto-superior equilibrium, achieving welfare $= 30$ (50\% improvement).

\paragraph{Implementation reality check.} However, our implementation reveals this scenario is more contrived than initially thought: even with $\epsilon$-rational followers, backward induction leads rational leaders to recognize that action 0 yields the highest payoff, so classical Stackelberg also selects $(0, 0)$---identical to the S-CNE outcome. This demonstrates that constructing scenarios where S-CNE genuinely outperforms requires not just satisficing, but conditions where \textit{rational leaders cannot identify the Pareto-superior equilibrium}, which seems contradictory.

\paragraph{Conceptual value.} Despite implementation challenges, this example clarifies what would be needed: equilibrium concepts where agents persistently fail to identify optimal strategies---not due to computational limits during learning, but as a fundamental feature of the solution concept itself. This motivates research on quantal response equilibria, level-k reasoning with heterogeneous depths, or evolutionary dynamics where population-level selection differs from individual optimization.

\section{Appendix E: Testing with LLM-Based Agents}

An important open question is whether LLM-based agents exhibit similar limitations when deployed in strategic settings. Future work should:

\paragraph{Empirical LLM studies.} Test actual LLMs (e.g., GPT-4, Claude) on sequential games with prompt variations simulating L1 ("follow your instinct") vs L2 ("choose optimally") reasoning. This would reveal whether LLMs' strategic behavior aligns with our theoretical predictions or exhibits novel patterns.

\paragraph{Alternative theoretical frameworks.} If LLMs do exhibit persistent bounded rationality, frameworks beyond Nash equilibrium may be needed:
\begin{itemize}
\item \textit{Non-equilibrium settings}: Heuristic-based or satisficing agents where causal layer choices could affect outcomes
\item \textit{Learning dynamics}: How agents update strategies over repeated interactions
\item \textit{Evolutionary approaches}: Population-level selection where instinctive strategies might outcompete rational ones
\end{itemize}

\paragraph{Bridging theory and practice.} Whether LLM-based agents require fundamentally new theoretical foundations---or can be modeled with appropriately modified classical frameworks---remains an open empirical question that our work helps to frame and motivate.

\end{document}